\documentclass[%
prd%
 ,twocolumn%
 ,secnumarabic%
,amssymb, amsmath,nobibnotes, aps, prl]{revtex4}
\usepackage{color}
\usepackage{graphicx}
\usepackage{multirow}

\begin{document}
\title{Influence of Young's modulus temperature dependence on parametric instability in Advanced LIGO interferometer}
\author{S.E. Strigin}
\affiliation{M.V. Lomonosov Moscow State University, Faculty of Physics, Moscow 119991, Russia}
\email{strigin@phys.msu.ru}
\begin{abstract}

We discuss the influence of Young's modulus temperature dependence on the number of parametrically unstable modes  in a Fabry-Perot cavity of Advanced LIGO interferometer. Some unstable modes may be suppressed  by changing the mirror's temperature due to temperature dependence of Young's modulus. Varying the temperature of the mirrors we can change their frequencies of the elastic modes; it, in turn, can change the number of unstable modes, leading to nonlinear effect of parametric oscillatory instability. The determination of  the optimal values of the temperature variations for some elastic modes to reduce the number of unstable modes is fulfilled.  Both  new ''fine tuning'' supression method of parametric instability and radius of curvature(ROC) change method  in the next generation of gravitational wave detectors are discussed. The applications of this method in cryogenic detectors like LIGO Voyager  or Einstein Telescope are proposed. 

\end{abstract}

\maketitle

\section{Introduction}



Experiments on the detection of parametric instability(PI) were carried out both in small-scale optomechanical setup\cite{chen} and then in the Advanced LIGO interferometer\cite{eva}.

In \cite{bsv} it was demonstrated the  PI in the Fabry-Perot (FP) cavities in Advanced LIGO interferometer could take place due to the high optical power  $W$   inside the arm cavities. It causes a transfer of energy from optical main mode(TEM$_{00}$) with frequency $\omega_0$  both to the mirror's elastic mode   with frequency $\omega_m$  and to an additional optical mode with frequency $\omega_1$(Stokes mode). The transfer of energy and substantial decrease of the interferometer's sensitivity occur at the resonance condition $\omega_0-\omega_1=\omega_m$. The full analysis of PIs for the LIGO interferometer  has been realized in \cite{gras3,2007,gras1,gras2,gras4,gras5,gras6,2007_1}. 
 
The condition of PI\cite{bsv,hei} for parametric gain ${\cal R}$ is:
\begin{eqnarray} 
\label{equ:gain1}
{\cal R} \simeq \frac{16 \pi}{c} \frac{W}{\lambda m \phi} \frac{1}{\omega_m^2 T_{tot}}
\times \frac{\Lambda_1}{1+\frac{\Delta^2}{\gamma_1^2}} \ ,
\\
\label{equ:ovlapfactor} \Lambda_1=\frac{V(\int A_0(r_{\bot})A_1(r_{\bot})u_z dr_{\bot})^2}{\int|A_0|^2
dr_{\bot}\int|A_1|^2dr_{\bot}\int|\vec{u}|^2 dV} \ ,
\end{eqnarray}
where   $c$ is the speed of light,  $m$ is the mirror's mass, $\phi$ is a value of material loss angle, $W$ is the power inside the arm cavity, $\lambda$ is the wavelength of the optical main mode, $T_{tot}$ is an effective transmission coefficient\cite{hei},   $\gamma_1$ is the relaxation rate of the Stokes mode and $\Delta=\omega_0-\omega_1-\omega_m$ is the detuning value. The parameter $\Lambda_1$ is the overlapping factor between the elastic and the two optical modes (main mode and Stokes mode), $A_0$ and $A_1$ are the optical field distributions at the  surface of the mirror for the optical main mode and the Stokes mode. The vector $\vec{u}$ is the spatial displacement of the elastic mode and $u_z$ is the z-component of
$\vec{u}$  along the cylindrical axis. $\int d{r}_{\bot}$ corresponds to the integration over this reflecting surface and $\int dV$ -- over the mirror volume $V$.
For the estimation of all elastic mode shapes and frequencies we use finite element modelling in package COMSOL$^\circledR$ with accuracy estimates to be about 1\%. We consider Laguerre-Gauss  optical modes $LG_{nm}$ with transversal orders  $2n+m\leq 10$, where $n = 0, 1, 2,\ldots$ is the radial index and $m = 0, 1, 2,\ldots$ is the azimuthal index.

In this Letter we analyze the influence of Young's modulus temperature dependence on parametric instability in Advanced LIGO interferometer.
In section I we discuss all proposed different types of thermal methods for effective suppression of PI. In section II we estimate the dependence of the number of unstable modes on Young's modulus value for different material loss angles. It is worth noting that Young's modulus temperature dependence is ''more fine tuning'' in minimization of PIs in comparison with tuning of mirror's ROC by heating and has to be also considered in the next gravitational-wave detectors.

 \subsection{ I. THERMAL CORRECTION SYSTEMS}
 
In recent years, many methods of heating have been proposed to avoid the PI. It is well-known that the absorption of a laser power\cite{1512.08813_2} in the mirrors generates a temperature gradient of a few Kelvin across each test mass\cite{1512.08813_4}. In turn, it creates a thermal lens in a substrate\cite{1512.08813_5}. Several thermal compensation systems\cite{1512.08813_4,Lawrence}  were proposed to place in the detectors in order to mitigate these aberrations. Different types of  ring heaters\cite{1512.08813_13,1512.08813_15}   heat the outer edge of each test mass to reduce the temperature gradient in the test masses\cite{Lawrence,1512.08813_17,1512.08813_18}. In Advanced LIGO the ring heater have a position around the barrel of the test mass near one end and radiates power onto the test mass in the infrared region that is absorbed efficiently by the fused silica test mass surface.

The   Thermal Correction System(TCS)\cite{T1500290,1512.08813_2,1608.02934} in Advanced LIGO is required to optimize the spatial modes inside the interferometer due to their  degradation by imperfections in the mirrors caused
by radii of curvature or surface figure errors. The TCS  can alter the test mass ROC and, in turn,  adjust the transverse optical mode spacing in the interferometer. Therefore, it potentially controls parametric instabilities at high circulating optical power.  The TCS consists of three elements: a radiative ring heater(RH), a $CO_2$ laser projector and a Hartmann wavefront sensor (HWS). The usage of RHs  optimizes the radii of curvatures of the input test masses(ITMs) and end test masses(ETMs), in turn,
the $CO_2$ laser compensates the thermal lens in the ITM  by acting on the compensation plate\cite{T1500290_41}. The compensation plate is suspended behind the input mirror to compensate the thermal effects. The RH has to generate a homogeneous heating profile with minimal heating of the suspension structure.   The $CO_2$ system is uncoupled from the ring heater\cite{THESIS-DOCTOR,1512.08813_13,Che,Che_38}. The HWS is used to measure the test mass thermal aberrations.
For example, in \cite{Lawrence}  the possible mean variation of temperature in the substrate lies in the range $10-30^\circ\!C$ for different mirror's materials.

Other  way to reduce the heating and to  develop  the RHs which produce more symmetric heating profiles\cite{T1500290_41} is to coat the barrel of the mirror with a thin layer (a few microns) or an infrared reflecting metal such as a gold\cite{T1500290_41}. This  gold barrel coating reduces the radial heat flow, has a small influence on thermal noise and, in turn,  homogenizes the temperature distribution inside the substrate.   Gold coating applied to the barrel for thermal
compensation purposes may decrease  the elastic modes   quality factors enough  and reduce some  parametric instabilities.   This technique may be useful in a third generation detectors. For example, for Voyager blue design it is proposed to use black coating on the mirror barrel for radiative cooling\cite{Blair-Voyager}.

However, thermal tuning based on ring heaters with radiant heating(as in the 
Advanced LIGO detectors)  may not be sufficient in PI suppression  because of high elastic mode density. Tuning the cavity away from one unstable elastic mode   leads it to be tuned into resonance with  other elastic mode.      Another proposed type of the  thermal compensation  by thermal modulation of
the end mirror, combined with back side static heating, can give sufficient parametric
instability suppression\cite{1702.01899}. If  end mirror is heated  due to absorption of optical power (may be, by the main cavity mode or by a separate $CO_2$ laser beam applied to end mirror), thermal distortion of the mirror will change its ROC.   It is worth noting that for a power absorption of  20W one can estimate the mean temperature variation of the mirror to be about 10K\cite{1702.01899}. It implies that the $CO_2$ laser heating does not have a significant effect on the thermal noise level.

Due to modulated heating power(with a constant 1W heat power on a front surface) there would be  a static thermal deformation of the mirror. To overcome
this mirror deformation, it is applied an additional ring type heating beam(17W power) to the back surface of the test mass
to compensate the deformation created by   the front surface heating. It is worth noting heating of the test mass by a $CO_2$ laser requires high stability of the $CO_2$ laser power(to avoid noise coupling).

\subsection{ II. RESULTS}
Well-known possible method to minimize the number of unstable modes is the change of mirrors' radii of curvature by local heating using thermal lensing by ring heaters. This idea was firstly pointed out in \cite{gras3,gras1,gras7} for LIGO interferometer. The thermal tuning of optical FP cavities has been proposed to reduce the parametric gain of parametric instabilities\cite{gras7}. The authors investigated the performance achievable for such tuning obtained by thermal actuation of the mirrors of the arm cavities. 

\begin{figure}[t]
	\begin{center}
	\includegraphics[width=6cm,totalheight=5cm]{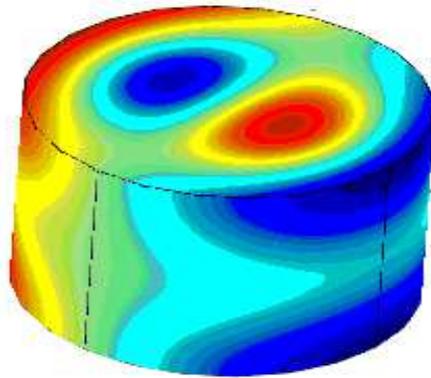}
		\caption{ The unstable elastic mode with frequency 15546Hz  and parametric gain  ${\cal R}_{\cal II}=1.14$ for material loss angle $\phi_{\cal II} = 2\times 10^{-8}$.  }	
	\label{fig:15546}
	\end{center}
\end{figure}

On the one hand, the small homogeneous heating(or  cooling) of the whole mirror can decide the problem of a minimization of the number of unstable elastic modes due to the  temperature dependence of Young's modulus value. On the other hand, these two effects of PI suppression(either change of ROC or Young's modulus by heating)  have to be taken into account simultaneously. It is necessary to combine these methods in order to less heat and change the radius of curvature of the  mirrors. In our Letter  we propose to do small modifications because of making small changes in Advanced LIGO design.

  \begin{table}
\begin{tabular} {|c|c|c|c|c|c} \hline
$f_m$[Hz] & Stokes mode & ${\cal R_I}$  &  ${\cal R_{II}}$ & ${\cal R_{III}}$\\\hline
15031 & $LG_{11}$& 2.66 & 1.33& 1.06   \\\hline
15546 & $LG_{11}$& 2.28 & 1.14 & 0.91 \\\hline
20165 & $LG_{20}$& 19.48 & 9.74& 7.79  \\\hline
25086 & $LG_{21}$& 3.35 & 1.68 & 1.34 \\\hline
29990 & $LG_{30}$& 6.5 & 3.25 & 2.6 \\\hline
30310 & $LG_{14}$& 1.66 & 0.83 & 0.66 \\\hline
30421 & $LG_{22}$& 5.05 & 2.5 & 2 \\\hline
31021 & $LG_{30}$& 6.6 & 3.3 & 2.64 \\\hline
\end{tabular}
\caption{ The values of parametric gains ${\cal R_I}$, ${\cal R_{II}}$ and  ${\cal R_{III}}$ for material loss angle $\phi_{\cal I} = 1\times 10^{-8}$, $\phi_{\cal II} = 2\times 10^{-8}$ and $\phi_{\cal III} = 2.5\times 10^{-8}$ correspondingly.}
\label{unstable_modes}
\end{table}

In Table \ref{unstable_modes} all possible unstable   elastic modes for three values of loss angle in Advanced LIGO mirrors are shown.  The    unstable elastic mode $15546{\rm Hz}$(see Fig.\ref{fig:15546}) with parametric gain ${\cal R_{II}}=1.14$ becomes stable if the variation of mirror's temperature is about $\Delta T \simeq 19^{\circ}C$(see Table \ref{unstable_modes_temperature}). However, if we heat the mirror by less than 15 degrees, the mode 15546Hz will become more stable, and, in turn, mode 30310Hz  will not become unstable. On the other hand, the elastic mode $15031{\rm Hz}$ requires the mirror's cooling with temperature variation $\Delta T \simeq -30{\rm}^{\circ}C$ to become absolutely stable. Other temperature variations for some unstable modes to have parametric gain equal to unity are shown in Table \ref{unstable_modes_temperature}.  It is worth noting that some unstable elastic modes can not become stable even at any possible temperature variations.
 There are 7 unstable elastic modes if loss angle $\phi_{\cal II} = 2\times 10^{-8}$. 
 If we cool the mirror by $30^\circ\!C$, the  mode 15031Hz will become stable, the mode 25086Hz will become less unstable and the mode 30310Hz --  more stable. The mode 15546Hz will be more unstable. For $\phi_{\cal II} = 2.5\times 10^{-8}$  there are 6 unstable elastic modes. If we decrease the mirror's temperature by no more than $\Delta T=13^\circ\!C$, the  mode 15031Hz will become stable, the mode 25086Hz -- less unstable. The mode 15546Hz remains   unstable. 

In all our preliminary calculations we do not take into account the small ROC change of the mirrors due to their  temperature variations. Therefore, the real temperature variations to reach the threshold value of parametric gain(${\cal R}=1$) may have slightly different values. In general,  this  Young's modulus temperature dependence has to be taken into account simultaneously with ROC change method to effectively  suppress the parametric instabilities.

 \begin{table}
\begin{tabular} {|c|c|c|c| } \hline 
$f_m$[Hz] & Stokes mode & $\Delta T_{\cal II},^\circ\!C$ & $\Delta T_{\cal III },^\circ\!C$ \\\hline
15031 & $LG_{11}$& -30 & -6   \\\hline
15546 & $LG_{11}$& 19 & -13   \\\hline
20165 & $LG_{20}$&  always unstable& always  unstable   \\\hline
25086 & $LG_{21}$& -50   &  -29 \\\hline
29990 & $LG_{30}$& always unstable & always unstable   \\\hline
30310 & $LG_{14}$& 15  &  36  \\\hline
30421 & $LG_{22}$& always unstable &  always unstable   \\\hline
31021 & $LG_{30}$& always unstable & always unstable  \\\hline
\end{tabular}
\caption{Mirrors' temperature variations   to reach the threshold value of parametric gain ${\cal R}=1$ for material loss angles   $\phi_{\cal II} = 2\times 10^{-8}$ and $\phi_{\cal III} = 2.5\times 10^{-8}$  in the temperature range from $-50^\circ\!C$ to $+50^\circ\!C$ correspondingly.  }
\label{unstable_modes_temperature}
\end{table}

\section{Conclusion}
In this Letter we have analyzed the influence of Young's modulus temperature dependence on the number of parametrically unstable modes  in a Fabry-Perot cavity of Advanced LIGO interferometer. It is proposed to use the possibility of the reduction of the unstable modes in Advanced LIGO by varying the temperature value in addition to changing the ROC of the mirrors. The frequencies of the elastic modes of the Fabry-Perot mirrors  change in this case. It can change the number of unstable modes leading to parametric oscillatory instability.


Our results of PI suppression are only preliminary estimates. We  take into account only Young's modulus temperature dependence because the temperature dependence of density and Poison ratio are negligible ones to be compared with Young's modulus dependence. On the other hand, the change in mirror's ROC by heating is more effective method for suppression of PIs. But Young's modulus temperature dependence is ''more fine tuning'' and  has to be  considered in the next gravitational-wave detectors like LIGO Voyager or Einstein Telescope.

It is known that future LIGO Voyager(blue design)  detector is planned to operate  at 123K temperature due to zero thermal expansion coefficient in this temperature  region. Tuning of the ROC thermally may not be effective for cryogenic detectors\cite{Blair-Voyager}.  However, since there are fewer unstable modes(two unstable modes with maximum parametric gain of 76) then in Advanced LIGO, it would be relatively easy to use other methods such as electrostatic feedback or passive damper to suppress PI. Therefore, combination of all method(thermal tuning, electrostatic feedback and passive dampers\cite{miller,eva}) will be able to eliminate
all instabilities. At the same time, Young's modulus temperature dependence method  may also give  good chance to avoid some unstable modes. The same situation will be in ET.

It also should be noted that either small homogeneous heating or cooling of the mirrors is a possible solution to suppress a more dangerous unstable elastic mode with a maximum parametric gain value(to derive it from resonance). A slight change in the mirror temperature, which differs from the approved temperature regime of the cavity mirrors, has to be done in the detectors to suppress some PIs. However, thermal methods of the PI reduction (changes in the ROCs of the mirrors and changes in the Young's modulus due to temperature)  must be considered simultaneously because of their dependence on each other. For example, for  high finesse case in LIGO Voyager(blue design)  there are several windows for different ROCs with only one unstable mode\cite{Blair-Voyager} which will be easier to suppress.  To suppress only one unstable mode in cryogenic detector  LIGO Voyager, it may be better to use the temperature dependence of Young's modulus. Small temperature variations(heating or cooling of the mirrors) make it possible to select another point around 123 K for the operating mode of the interferometer giving an insignificant increase in the level of thermal noise. These methods do not require significant changes in existing models of the interferometers.


Here it is worth noting two cases. Firstly, for cryogenic detectors it is possible to choose the optimal value of the ROC of the mirrors with the minimum number of unstable modes, and then apply the method of the temperature dependence of Young's modulus to suppress remaining(one or two) PIs. In turn, small homogeneous cooling or heating   of the whole  mirror  will not greatly change the radius of curvature. Secondly, for room temperature detectors PI suppression  due to  coating of the mirror barrel  with a thin layer(as a passive method to decrease mechanical quality factors) and homogeneous heating of the substrate by ring heater simultaneously is  very promising. At the same time, in cryogenic detectors tuning of the ROC by ring heater is undesirable operation. In this case   coating of the mirror barrel with a thin layer(as a passive method) and then, for the remaining one or two unstable modes, the optimal choice of the cryogenic temperature of mirrors(using Young's modulus dependence) will give effective suppression of PIs. At present this work is in progress for LIGO Voyager and ET.

\section*{Acknowledgement}
The author kindly acknowledges the support  of the Russian Science Foundation (grant 17-12-01095).  

\begin{appendix}

\section{A. Parameters   of the calculations}
\label{app:params}
In this Appendix we present the numerical values of the parameters of Advanced LIGO interferometer  that influence our calculations. 
As the substrates are made from fused silica we used material parameters according to this material at room temperature. 
The numerical values are presented in Table~\ref{tab:props1}.  

\begin{table}
\begin{tabular}{|lc|}
\multicolumn{2}{l}{elastic modes}\\
\hline
ETM radius 								& 17\,cm \\
ETM height									&	20\,cm \\
Young's modulus $Y$					&73.1\,GPa \\
Poisson's ratio $\sigma$		&0.17 \\
density $\rho$							&2203\,kg/m$^3$ \\
\hline
\multicolumn{2}{l}{optical modes} \\
\hline
\multicolumn{2}{|l|}{radius of curvature}							\\
ITM	$R_1$												&1934\,m \\
ETM	$R_2$												&2245\,m \\
\hline
\multicolumn{2}{|l|}{substrate diameter}	\\	
ITM													&34\,cm \\
ETM													&34\,cm		\\
\hline
\multicolumn{2}{|l|}{laser beam radius $w$} \\
ITM													&5.3\,cm \\
ETM													&6.2\,cm 	\\
\hline
cavity length $L$						&	3994.5\,m \\
\hline
\multicolumn{2}{l}{parametric gain}\\
\hline
\multicolumn{2}{|l|}{energy transmission}		\\
ITM	$T_\mathrm{ITM}$				&	$14\times10^{-3}$			\\
ETM													& 0	\\
\hline
mechanical loss $\phi$				&$1\times10^{-8} -  2.5\times10^{-8}$ \\
laser wavelength $\lambda$	& 1064\,nm \\
mirror mass (ETM) $m$						&40 \,kg \\
optical power inside cavity $W$	&0.83\,MW \\
speed of light $c$					&$3\times 10^{8}$\,m/s \\
\hline
\end{tabular}
\caption{Numerical values of the parameters for the calculation of parametric instabilities.}
\label{tab:props1}
\end{table}


\end{appendix}

\bibliographystyle{iopart-num}
\bibliography{refs}

\end{document}